\documentclass{appolb}
\usepackage{graphicx}
\usepackage{amsmath}

\begin{document}
\eqsec  
\title{Meson vacuum phenomenology in a three-flavor linear sigma model
  with (axial-)vector mesons: investigation of the $U(1)_A$ anomaly term
\thanks{Presented at Excited QCD 2013 (3-9 February 2013, Bjelasnica
  Mountain, Sarajevo)}
}
\author{P{\'e}ter Kov{\'a}cs\footnote{kovacs.peter@wigner.mta.hu} , Gy{\"o}rgy Wolf
\address{Institute for Particle and Nuclear Physics, Wigner Research
  Center for Physics, H-1525 Budapest, POB 49, Hungary}
}
\maketitle

\begin{abstract}
Zero temperature properties of an (axial-) vector meson extended
linear $\sigma$-model are discussed, concerning on the possible different
realizations of the axial anomaly term. The different anomaly terms
are compared with each other on the basis of a $\chi^2$ minimalization process. It is
found that there is no essential difference among the different
realizations. This means that any of them can be equally used from
phenomenological point of view. 
\end{abstract}

\PACS{12.39.Fe, 12.40.Yx, 14.40.Be, 14.40.Df}
  
\section{Introduction}

According to our knowledge the commonly believed fundamental theory of strong
interaction is Quantum Chromodynamics (QCD), which is up to now proved
to be unsolvable in the low energy regime, where the basic degrees
of freedom are the observable mesons and hadrons. Since the original
degrees of freedom in QCD are quarks and gluons and the construction
of mesons and hadrons is unknown, in this regime one possibility is to
build up some effective theory \cite{geffen}, which
reflects some of the original properties of QCD. One of the most
important such property is the approximate global $U(3)_{L}\times
U(3)_{R}$ symmetry (if we consider three flavours), the chiral
symmetry. This symmetry is isomorphic to the $U(1)_V\times SU(3)_V
\times U(1)_A\times SU(3)_A$, which is broken down -- explicitly due
to nonzero quark masses, and spontaneously due to nonzero quark
condensates \cite{ssb} -- to $U(1)_V\times SU(3)_V$, if the isospin
symmetric case is considered.

Since if only the chiral symmetry is considered alone, the Lagrangian
can still contain infinitely many terms, it is necessary to impose
other restrictions as well. One natural choice is renormalizability,
however in an effective model this is not totally necessary. Moreover,
since we would also like to include (axial-) vector mesons, the
renormalizability is violated anyway. Instead of renormalizability we
have chosen dilaton symmetry to restrict the number of terms (for more
details see \cite{article1} and references therein).

In order to maintain the above mentioned symmetry breaking pattern
($U(1)_V \times SU(3)_V \times U(1)_A\times SU(3)_A \longrightarrow
U(1)_V \times SU(3)_V$), besides the chiral and dilaton symmetric
terms, symmetry braking terms are also needed in the effective
Lagrangian. The symmetry is broken explicitly and spontaneously, and
concerning on the $U(1)_A$ violation the spontaneous braking is
realized through the so-called axial/chiral anomaly
\cite{anomaly}. Because of this, an anomaly term should be introduced
into the effective Lagrangian, which can have different forms, as will
be discussed shortly. 

The specific form of the anomaly term will affect the form of the
tree level masses of the pseudoscalar meson sector. Thus, through a
$\chi^2$ minimalization process, which maintains a comparison between the
model predictions and the physical spectrum, the 'goodness' of the
different anomaly terms can be investigated.  

The paper is organized as follows. In Sec. 2 we briefly discuss the
model -- which is described in more detail in our previous works
\cite{articles2011,articles2012,article1}. In Sec. 3 we investigate
the different anomaly terms and describe the $\chi^2$ minimalization
process, while in Sec. 4 we conclude.

\section{The model}

For the Lagrangian we use, apart from a modified anomaly term
$\mathcal{L}_{U(1)_A}$, the same as in \cite{article1}, where we have
neglected the dilaton field, since it is irrelevant in the current
investigation. Thus, our Lagrangian takes the following form, 
\begin{align}
\mathcal{L}  & = \mathop{\mathrm{Tr}}[(D_{\mu}\Phi)^{\dagger}(D_{\mu}\Phi)]-m_{0}^{2}
\mathop{\mathrm{Tr}}(\Phi^{\dagger}\Phi
)-\lambda_{1}[\mathop{\mathrm{Tr}}(\Phi^{\dagger}\Phi)]^{2}-\lambda
_{2}\mathop{\mathrm{Tr}}(\Phi^{\dagger}\Phi)^{2}{\nonumber}\\
&  -\frac{1}{4}\mathop{\mathrm{Tr}}(L_{\mu\nu}^{2}+R_{\mu\nu}^{2}%
)+\mathop{\mathrm{Tr}}\left[  \left( \frac{m_{1}^{2}}{2}+\Delta\right)
  (L_{\mu}^{2}+R_{\mu}^{2})\right]
+\mathop{\mathrm{Tr}}[H(\Phi+\Phi^{\dagger})]{\nonumber}\\
&  +\mathcal{L}_{U(1)_A}+i\frac{g_{2}}{2}%
(\mathop{\mathrm{Tr}}\{L_{\mu\nu}[L^{\mu},L^{\nu}%
]\}+\mathop{\mathrm{Tr}}\{R_{\mu\nu}[R^{\mu},R^{\nu}]\}){\nonumber}\\
&  +\frac{h_{1}}{2}\mathop{\mathrm{Tr}}(\Phi^{\dagger}\Phi
)\mathop{\mathrm{Tr}}(L_{\mu}^{2}+R_{\mu}^{2})+h_{2}%
\mathop{\mathrm{Tr}}[\vert L_{\mu}\Phi \vert ^{2}+\vert \Phi R_{\mu} \vert ^{2}]+2h_{3}%
\mathop{\mathrm{Tr}}(L_{\mu}\Phi R^{\mu}\Phi^{\dagger}){\nonumber}\\
&  +g_{3}[\mathop{\mathrm{Tr}}(L_{\mu}L_{\nu}L^{\mu}L^{\nu}%
)+\mathop{\mathrm{Tr}}(R_{\mu}R_{\nu}R^{\mu}R^{\nu})]+g_{4}%
[\mathop{\mathrm{Tr}}\left(  L_{\mu}L^{\mu}L_{\nu}L^{\nu}\right)
+\mathop{\mathrm{Tr}}\left(  R_{\mu}R^{\mu}R_{\nu}R^{\nu}\right)
]{\nonumber}\\
&  +g_{5}\mathop{\mathrm{Tr}}\left(  L_{\mu}L^{\mu}\right)
\,\mathop{\mathrm{Tr}}\left(  R_{\nu}R^{\nu}\right)  +g_{6}%
[\mathop{\mathrm{Tr}}(L_{\mu}L^{\mu})\,\mathop{\mathrm{Tr}}(L_{\nu}L^{\nu
})+\mathop{\mathrm{Tr}}(R_{\mu}R^{\mu})\,\mathop{\mathrm{Tr}}(R_{\nu}R^{\nu
})]\text{ ,} \label{eq:Lagrangian}%
\end{align}
where  
\begin{align}
D^{\mu}\Phi &  \equiv\partial^{\mu}\Phi-ig_{1}(L^{\mu}\Phi-\Phi R^{\mu
})-ieA^{\mu}[T_{3},\Phi]\;,\nonumber\\
L^{\mu\nu}  &  \equiv\partial^{\mu}L^{\nu}-ieA^{\mu}[T_{3},L^{\nu}]-\left\{
\partial^{\nu}L^{\mu}-ieA^{\nu}[T_{3},L^{\mu}]\right\}  \;\text{,}\nonumber\\
R^{\mu\nu}  &  \equiv\partial^{\mu}R^{\nu}-ieA^{\mu}[T_{3},R^{\nu}]-\left\{
\partial^{\nu}R^{\mu}-ieA^{\nu}[T_{3},R^{\mu}]\right\}  \;\text{.}\nonumber
\end{align}
The quantities $\Phi$, $R^{\mu}$, and $L^{\mu}$ represent the scalar and
vector nonets:
\begin{align}
\Phi &  =\sum_{i=0}^{8}(S_{i}+iP_{i})T_{i},\quad 
L^{\mu}  &  =\sum_{i=0}^{8}(V_{i}^{\mu}+A_{i}^{\mu})T_{i},\quad
R^{\mu}  &  =\sum_{i=0}^{8}(V_{i}^{\mu}-A_{i}^{\mu})T_{i}
\end{align}
where, $T_{i}\,(i=0,\ldots,8)$ denote the generators of $U(3)$, while
$S_{i}$ represents the scalar, $P_{i}$ the pseudoscalar, $V_{i}^{\mu}$ the
vector, $A_{i}^{\mu}$ the axial-vector meson fields, and $A^{\mu}$ is the
electromagnetic field. 

It is worth to note that in the $(0-8)$ sector there is a particle
mixing (see \cite{article1}) and we use the
$\varphi_{N}  = \frac{1}{\sqrt{3}}\left(  \sqrt{2}\;\varphi_{0}+\varphi
_{8}\right)$, 
$\varphi_{S}  =\frac{1}{\sqrt{3}}\left(  \varphi_{0}-\sqrt{2}\;\varphi
_{8}\right)$,  $\varphi\in(S_{i},P_{i},V_{i}^{\mu},A_{i}^{\mu})$ non strange --
strange basis, which is more suitable for our calculations. 
Moreover, $H$ and $\Delta$ are constant external fields defined as
$H = H_{0}T_{0} + H_{8}T_{8} = \mathrm{diag}\left(h_{0N}/2, h_{0N}/2,
  h_{0S}/\sqrt{2} \right)$, $\Delta  = \Delta_{0}T_{0} +
\Delta_{8}T_{8} = \mathrm{diag}\left(\delta_{N}, \delta_{N}, \delta_{S} \right)$

Finally for the $\mathcal{L}_{U(1)_A}$ anomaly term we use three
different terms,
\begin{align}
\mathcal{L}_{U(1)_A} & = c_1(\det\Phi+\det\Phi^{\dagger}) +
c_2(\det\Phi-\det\Phi^{\dagger})^2 \\
& + c_m(\det\Phi+\det\Phi^{\dagger})\Tr(\Phi\Phi^{\dagger}),
\end{align}
which should be understand in a sense that from the $c_1, c_2, c_m$
parameters only one is different from zero at the same time. The
first two terms are approximations of the ``original'' axial anomaly
term, which is $\propto (\ln\det\Phi-\ln\det\Phi^{\dagger})$ (see the
original term e. g. in \cite{rosenzweig1979} and the approximation in
\cite{fariborz2008}), while the third term is a mixed term. Our
concept was to choose different anomaly terms in which the power of the
$\Phi$ field is no more than six. This can be regarded as a first
approximation used to compare the effects of different anomaly terms
on the spectrum.

\section{Comparison of the different anomaly terms}

For the analysis we used a $\chi^2$ method (for more details see
\cite{article1}) in which we calculated
some physical quantities -- masses and decay widths -- at tree-level,
and compared to experimental data taken from the PDG
\cite{PDG}. $\chi^2$ is defined as
\begin{equation}
  \chi^2(x_1,\dots,x_N)=\sum_{i=1}^{M}\left[\frac{Q_i(x_1,\dots,x_N)-Q_i^{\text{exp}}}
    {\delta Q_i}\right]^2,
\end{equation}
where $(x_1,\dots,x_N)=(m_0, \lambda_1, \lambda_2,\dots)$ are the
unknown parameters of the model, $Q_i(x_1,\dots,x_N)$ are the
calculated physical quantities, while $Q_i^{\text{exp}}\pm \delta Q_i$
are the experimental values taken from the PDG. In the process we
minimalize the $\chi^2$ and determine the 11 unknown parameters of the
model, which are 
$C_{1} (\equiv m_{0}^{2} + \lambda_{1} \left(  \phi_{N}^{2}
  +\phi_{S}^{2}\right)), C_{2}(\equiv m_{1}^{2} + \frac{h_{1}}{2} \left(  \phi_{N}^{2} + \phi
_{S}^{2}\right)), \delta_{S}, g_{1}, g_{2},$ $\phi_{N},$ $\phi_{S},$
$h_{2},$ $h_{3},$ $\lambda_{2}$ and
one from $c_1, c_2, c_m$. The determined parameters belonging to the
minimal $\chi^2$ give the best description of the experimental
data. For the physical quantities we have chosen the following 21
observables \cite{article1}, $f_{\pi }$, $f_{K} $, $m_{\pi }$,
$m_{K}$, $m_{\eta }$, $m_{\eta ^{\prime }}$, $m_{\rho }$
, $m_{K^{\star }}$, $m_{\omega _{S}\equiv \varphi (1020)}$, $m_{f_{1S}\equiv
f_{1}(1420)}$, $m_{a_{1}}$, $m_{a_{0}\equiv a_{0}(1450)}$, $m_{K_{0}^{\star
}\equiv K_{0}^{\star }(1430)}$, $\Gamma _{\rho \rightarrow \pi \pi }$,
$\Gamma _{K^{\star }\rightarrow K\pi }$, $\Gamma _{\phi \rightarrow
  KK}$, $\Gamma_{a_{1}\rightarrow \rho \pi }$, $\Gamma
_{a_{1}\rightarrow \pi \gamma}$, $\Gamma _{f_{1}(1420)\rightarrow
  K^{\star }K}$, $\Gamma _{a_{0}(1450)}$, $\Gamma _{K_{0}^{\star
  }(1430)\rightarrow K\pi }$.\footnote{It is worth to note that according to
the different anomaly terms the functional form of some of the
observables are different that in \cite{article1}.}

In Table \ref{tab1} we summarized the $\chi^{2}$ and the reduced
$\chi^{2}_{\text{red}}$ values for the different cases. In the last row
$c_{\text{all}}$ means that we included all the three
anomaly terms in the fit. 
\begin{table}[th]
\centering
\begin{tabular}
[c]{|c|c|c|}\hline
term & $\chi^{2}$ & $\chi_{\text{red}}^{2}$\\\hline
$c_1$ & $59.38$ & $5.94$\\\hline
$c_2$ & $62.40$ & $6.24$\\\hline
$c_m$ & $110.93$ & $11.09$\\\hline
$c_{\text{all}}$ & $50.19$ & $6.27$\\\hline
\end{tabular}
\caption{The total $\chi^{2}$ and the reduced
  $\chi^{2}_{\text{red}}=\chi^{2}/N_{\text{dof}}$ for the different anomaly terms,
where $N_{\text{dof}}$ is the difference between the number of experimental
quantities and the number of fit parameters (10 for the first three
row and 8 for the last).}
\label{tab1}
\end{table}
It can be seen that the $c_1$ and $c_2$ terms give basically the
same description of the experimental data, while the $c_m$ term is not
as good as the first two. Even if we include all the three terms and
extend the number of free parameters, we can not get any closer to
describe better the experimental data. We can therefore conclude that
in the linear sigma model one can use either the $c_1$ or the $c_2$ type of term
for the anomaly, while the use of some mixed term is not favored. A
detailed analysis shows that either the $c_1$ or $c_2$ term is in good
agreement with the experimental data.

\section{Conclusion}

We presented an (axial-) vector meson extended linear sigma model with
different axial anomaly terms. Global $\chi^2$ fits were performed in
order to compare the different anomaly terms. All the model parameters were
fixed in this $\chi^2$ minimalization process. For the different
anomaly terms we found that two of them -- which are emanating from
the originally suggested anomaly term -- describe basically the same
physics and are in good agreement with the experimental data, while
the third one, the mixed term, is not as good as the others. 

We should point out, however, that the presented investigation is only a small
part of the meson phenomenology, which aims to understand better the
mechanisms of strong interaction at low energies.

\section*{Acknowledgment}

The authors were supported by the Hungarian OTKA funds T71989 and
T101438. Furthermore the authors would like to thank for the colleges
F.\ Giacosa, D.\ Parganlija and D.H.\ Rischke for the useful discussions.


\end{document}